\documentclass[aps,prl,showpacs,twocolumn,superscriptaddress]{revtex4}  

\usepackage{graphicx}
\usepackage{dcolumn}
\usepackage{bm}
\usepackage{amssymb}   
\usepackage{epsfig}

\usepackage{bbm}
\usepackage{url}

\newcommand{\fb} {{\rm fb}^{-1}}

\newcommand{\lumi} {4.4~\fb}

\newcommand{\GeV}{{\rm GeV}}
\newcommand{\GeVc}{{\rm GeV/}c}
\newcommand{\GeVcsq}{{\rm GeV/}c^2}

\newcommand{\MeVcsq}{{\rm MeV/}c^2}

\newcommand{\kmmRelBr}{0.38}
\newcommand{\kmmRelBrStat}{0.05}
\newcommand{\kmmRelBrSyst}{0.02}
\newcommand{\kstmmRelBr}{0.80}
\newcommand{\kstmmRelBrStat}{0.10}
\newcommand{\kstmmRelBrSyst}{0.06}
\newcommand{\phimmRelBr}{1.11}
\newcommand{\phimmRelBrStat}{0.25}
\newcommand{\phimmRelBrSyst}{0.09}

\newcommand{\kmmBr}{0.38}
\newcommand{\kmmBrStat}{0.05}
\newcommand{\kmmBrSyst}{0.03}
\newcommand{\kstmmBr}{1.06}
\newcommand{\kstmmBrStat}{0.14}
\newcommand{\kstmmBrSyst}{0.09}
\newcommand{\phimmBr}{1.44}
\newcommand{\phimmBrStat}{0.33}
\newcommand{\phimmBrSyst}{0.46}

\newcommand{\bz}{B^0}

\newcommand{\bp}{B^+}

\newcommand{\bs}{B^0_s}

\newcommand {\kst}{K^{*0}}

\newcommand {\kstpdg}{K^{*}(892)^{0}}

\newcommand {\phipdg}{\phi(1020)}

\newcommand{\kmm}{K^+ \mu^+ \mu^-}

\newcommand{\kstmm}{\kst \mu^+ \mu^-}

\newcommand{\phimm}{\phi \mu^+ \mu^-}

\newcommand{\hmm}{h \mu^+ \mu^-}

\newcommand{\bsll}{b \to s \ell \ell}

\newcommand{\kstll}{\kst \ell^+ \ell^-}

\newcommand{\smm}{s \mu \mu}
\newcommand{\bsmm}{b \to \smm}

\newcommand{\jpsi}{J/{\small \psi}}
\newcommand{\psip}{\psi'}

\newcommand{\psih}{\jpsi h}

\newcommand{\bpkmm}{\bp \to \kmm}

\newcommand{\bzkstmm}{\bz \to \kstmm}

\newcommand{\bsphimm}{\bs \to \phimm}

\newcommand{\bhmm}{B \to \hmm}

\newcommand{\bzkstmmpdg}{\bz \to \kstpdg \mu^+ \mu^-}

\newcommand{\bsphimmpdg}{\bs \to \phipdg \mu^+ \mu^-}

\newcommand{\bzkstll}{\bz \to \kstll}

\newcommand{\bpsih}{B \to \psih}

\newcommand{\psimm}{\jpsi \to \mu^+ \mu^-}

\newcommand{\fl}{\rm F_{L}}
\newcommand{\afb}{\rm A_{FB}}

\newcommand{\cosm}{\cos \theta_\mu}
\newcommand{\cosk}{\cos \theta_K}
\newcommand{\cosmsq}{\cos^2 \theta_\mu}
\newcommand{\cosksq}{\cos^2 \theta_K}

\newcommand{\pythia}{{\sc pythia} }
\newcommand{\evtgen}{{\sc EvtGen}}

\newcommand{\br}{{\rm BR}}

\begin{document}


\title{\bf Measurement of the Forward-Backward Asymmetry \\
in the $B\to K^{(*)}\mu^+\mu^-$ Decay \\
and First Observation of the $\bsphimm$ Decay
}

\affiliation{Institute of Physics, Academia Sinica, Taipei, Taiwan 11529, Republic of China} 
\affiliation{Argonne National Laboratory, Argonne, Illinois 60439, USA} 
\affiliation{University of Athens, 157 71 Athens, Greece} 
\affiliation{Institut de Fisica d'Altes Energies, Universitat Autonoma de Barcelona, E-08193, Bellaterra (Barcelona), Spain} 
\affiliation{Baylor University, Waco, Texas 76798, USA} 
\affiliation{Istituto Nazionale di Fisica Nucleare Bologna, $^{bb}$University of Bologna, I-40127 Bologna, Italy} 
\affiliation{Brandeis University, Waltham, Massachusetts 02254, USA} 
\affiliation{University of California, Davis, Davis, California 95616, USA} 
\affiliation{University of California, Los Angeles, Los Angeles, California 90024, USA} 
\affiliation{Instituto de Fisica de Cantabria, CSIC-University of Cantabria, 39005 Santander, Spain} 
\affiliation{Carnegie Mellon University, Pittsburgh, Pennsylvania 15213, USA} 
\affiliation{Enrico Fermi Institute, University of Chicago, Chicago, Illinois 60637, USA}
\affiliation{Comenius University, 842 48 Bratislava, Slovakia; Institute of Experimental Physics, 040 01 Kosice, Slovakia} 
\affiliation{Joint Institute for Nuclear Research, RU-141980 Dubna, Russia} 
\affiliation{Duke University, Durham, North Carolina 27708, USA} 
\affiliation{Fermi National Accelerator Laboratory, Batavia, Illinois 60510, USA} 
\affiliation{University of Florida, Gainesville, Florida 32611, USA} 
\affiliation{Laboratori Nazionali di Frascati, Istituto Nazionale di Fisica Nucleare, I-00044 Frascati, Italy} 
\affiliation{University of Geneva, CH-1211 Geneva 4, Switzerland} 
\affiliation{Glasgow University, Glasgow G12 8QQ, United Kingdom} 
\affiliation{Harvard University, Cambridge, Massachusetts 02138, USA} 
\affiliation{Division of High Energy Physics, Department of Physics, University of Helsinki and Helsinki Institute of Physics, FIN-00014, Helsinki, Finland} 
\affiliation{University of Illinois, Urbana, Illinois 61801, USA} 
\affiliation{The Johns Hopkins University, Baltimore, Maryland 21218, USA} 
\affiliation{Institut f\"{u}r Experimentelle Kernphysik, Karlsruhe Institute of Technology, D-76131 Karlsruhe, Germany} 
\affiliation{Center for High Energy Physics: Kyungpook National University, Daegu 702-701, Korea; Seoul National University, Seoul 151-742, Korea; Sungkyunkwan University, Suwon 440-746, Korea; Korea Institute of Science and Technology Information, Daejeon 305-806, Korea; Chonnam National University, Gwangju 500-757, Korea; Chonbuk National University, Jeonju 561-756, Korea} 
\affiliation{Ernest Orlando Lawrence Berkeley National Laboratory, Berkeley, California 94720, USA} 
\affiliation{University of Liverpool, Liverpool L69 7ZE, United Kingdom} 
\affiliation{University College London, London WC1E 6BT, United Kingdom} 
\affiliation{Centro de Investigaciones Energeticas Medioambientales y Tecnologicas, E-28040 Madrid, Spain} 
\affiliation{Massachusetts Institute of Technology, Cambridge, Massachusetts 02139, USA} 
\affiliation{Institute of Particle Physics: McGill University, Montr\'{e}al, Qu\'{e}bec, Canada H3A~2T8; Simon Fraser University, Burnaby, British Columbia, Canada V5A~1S6; University of Toronto, Toronto, Ontario, Canada M5S~1A7; and TRIUMF, Vancouver, British Columbia, Canada V6T~2A3} 
\affiliation{University of Michigan, Ann Arbor, Michigan 48109, USA} 
\affiliation{Michigan State University, East Lansing, Michigan 48824, USA}
\affiliation{Institution for Theoretical and Experimental Physics, ITEP, Moscow 117259, Russia}
\affiliation{University of New Mexico, Albuquerque, New Mexico 87131, USA} 
\affiliation{Northwestern University, Evanston, Illinois 60208, USA} 
\affiliation{The Ohio State University, Columbus, Ohio 43210, USA} 
\affiliation{Okayama University, Okayama 700-8530, Japan} 
\affiliation{Osaka City University, Osaka 588, Japan} 
\affiliation{University of Oxford, Oxford OX1 3RH, United Kingdom} 
\affiliation{Istituto Nazionale di Fisica Nucleare, Sezione di Padova-Trento, $^{cc}$University of Padova, I-35131 Padova, Italy} 
\affiliation{LPNHE, Universite Pierre et Marie Curie/IN2P3-CNRS, UMR7585, Paris, F-75252 France} 
\affiliation{University of Pennsylvania, Philadelphia, Pennsylvania 19104, USA}
\affiliation{Istituto Nazionale di Fisica Nucleare Pisa, $^{dd}$University of Pisa, $^{ee}$University of Siena and $^{ff}$Scuola Normale Superiore, I-56127 Pisa, Italy} 
\affiliation{University of Pittsburgh, Pittsburgh, Pennsylvania 15260, USA} 
\affiliation{Purdue University, West Lafayette, Indiana 47907, USA} 
\affiliation{University of Rochester, Rochester, New York 14627, USA} 
\affiliation{The Rockefeller University, New York, New York 10065, USA} 
\affiliation{Istituto Nazionale di Fisica Nucleare, Sezione di Roma 1, $^{gg}$Sapienza Universit\`{a} di Roma, I-00185 Roma, Italy} 

\affiliation{Rutgers University, Piscataway, New Jersey 08855, USA} 
\affiliation{Texas A\&M University, College Station, Texas 77843, USA} 
\affiliation{Istituto Nazionale di Fisica Nucleare Trieste/Udine, I-34100 Trieste, $^{hh}$University of Trieste/Udine, I-33100 Udine, Italy} 
\affiliation{University of Tsukuba, Tsukuba, Ibaraki 305, Japan} 
\affiliation{Tufts University, Medford, Massachusetts 02155, USA} 
\affiliation{Waseda University, Tokyo 169, Japan} 
\affiliation{Wayne State University, Detroit, Michigan 48201, USA} 
\affiliation{University of Wisconsin, Madison, Wisconsin 53706, USA} 
\affiliation{Yale University, New Haven, Connecticut 06520, USA} 
\author{T.~Aaltonen}
\affiliation{Division of High Energy Physics, Department of Physics, University of Helsinki and Helsinki Institute of Physics, FIN-00014, Helsinki, Finland}
\author{B.~\'{A}lvarez~Gonz\'{a}lez$^v$}
\affiliation{Instituto de Fisica de Cantabria, CSIC-University of Cantabria, 39005 Santander, Spain}
\author{S.~Amerio}
\affiliation{Istituto Nazionale di Fisica Nucleare, Sezione di Padova-Trento, $^{cc}$University of Padova, I-35131 Padova, Italy} 

\author{D.~Amidei}
\affiliation{University of Michigan, Ann Arbor, Michigan 48109, USA}
\author{A.~Anastassov}
\affiliation{Northwestern University, Evanston, Illinois 60208, USA}
\author{A.~Annovi}
\affiliation{Laboratori Nazionali di Frascati, Istituto Nazionale di Fisica Nucleare, I-00044 Frascati, Italy}
\author{J.~Antos}
\affiliation{Comenius University, 842 48 Bratislava, Slovakia; Institute of Experimental Physics, 040 01 Kosice, Slovakia}
\author{G.~Apollinari}
\affiliation{Fermi National Accelerator Laboratory, Batavia, Illinois 60510, USA}
\author{J.A.~Appel}
\affiliation{Fermi National Accelerator Laboratory, Batavia, Illinois 60510, USA}
\author{A.~Apresyan}
\affiliation{Purdue University, West Lafayette, Indiana 47907, USA}
% ...
\collaboration{CDF Collaboration\footnote{With visitors from $^a$University of Massachusetts Amherst, Amherst, Massachusetts 01003,
$^b$Istituto Nazionale di Fisica Nucleare, Sezione di Cagliari, 09042 Monserrato (Cagliari), Italy,
$^c$University of California Irvine, Irvine, CA  92697, 
$^d$University of California Santa Barbara, Santa Barbara, CA 93106
$^e$University of California Santa Cruz, Santa Cruz, CA  95064,
$^f$CERN,CH-1211 Geneva, Switzerland,
$^g$Cornell University, Ithaca, NY  14853, 
$^h$University of Cyprus, Nicosia CY-1678, Cyprus, 
$^i$University College Dublin, Dublin 4, Ireland,
$^j$University of Fukui, Fukui City, Fukui Prefecture, Japan 910-0017,
$^k$Universidad Iberoamericana, Mexico D.F., Mexico,
$^l$Iowa State University, Ames, IA  50011,
$^m$University of Iowa, Iowa City, IA  52242,
$^n$Kinki University, Higashi-Osaka City, Japan 577-8502,
$^o$Kansas State University, Manhattan, KS 66506,
$^p$University of Manchester, Manchester M13 9PL, England,
$^q$Queen Mary, University of London, London, E1 4NS, England,
$^r$Muons, Inc., Batavia, IL 60510,
$^s$Nagasaki Institute of Applied Science, Nagasaki, Japan, 
$^t$National Research Nuclear University, Moscow, Russia,
$^u$University of Notre Dame, Notre Dame, IN 46556,
$^v$Universidad de Oviedo, E-33007 Oviedo, Spain, 
$^w$Texas Tech University, Lubbock, TX  79609, 
$^x$IFIC(CSIC-Universitat de Valencia), 56071 Valencia, Spain,
$^y$Universidad Tecnica Federico Santa Maria, 110v Valparaiso, Chile,
$^z$University of Virginia, Charlottesville, VA  22906,
$^{aa}$Yarmouk University, Irbid 211-63, Jordan,
$^{ii}$On leave from J.~Stefan Institute, Ljubljana, Slovenia, 
}}
\noaffiliation

\begin{abstract}

We reconstruct the rare decays $\bpkmm$, $\bzkstmmpdg$, and $\bsphimmpdg$
in a data sample corresponding to $\lumi$ 
collected in $p\bar{p}$ collisions at $\sqrt{s}=1.96~{\rm TeV}$
by the CDF II detector at the Tevatron Collider.
Using $121\pm16$ $\bpkmm$ and $101\pm12$ $\bzkstmm$ decays we report the branching ratios.
In addition, we report the differential branching ratio and the muon forward-backward asymmetry 
in the $\bp$ and $\bz$ decay modes, and the $\kst$ longitudinal
polarization fraction in the $\bz$ decay mode with respect to the squared dimuon mass.
These are consistent with the predictions,
and most recent determinations from other experiments and of comparable accuracy.
We also report the first observation of the $\bsphimm$ decay and measure its branching ratio $\br(\bsphimm) = [\phimmBr \pm \phimmBrStat \pm \phimmBrSyst] \times 10^{-6}$
using $27\pm6$ signal events. This is currently the most rare $\bs$ decay observed.

\end{abstract}

\maketitle

The flavor-changing neutral current process 
$\bsll$ occurs in the standard model (SM) through higher order diagrams 
where new physics contributions could arise.
Accurate SM predictions make the $\bsll$ phenomenology suited to uncover early indications of new physics~\cite{Ali:1999mm_APS,Kruger:2005ep_APS,Bobeth:2007dw_APS}, especially 
through observables like the lepton forward-backward asymmetry ($\afb$) and 
the differential branching fraction ($\br$) as a function of dilepton mass $M_{\ell\ell}$.
The $\bsll$ amplitudes can be described in terms of short distance operators and effective Wilson coefficients $C_{7,9,10}$.
Some new physics models~\cite{Ali:1999mm_APS} allow the flipped sign of $C_7$.
This results in the opposite sign of $\afb$ in the small $q^2$ region
($q^2\equiv M_{\ell\ell}^2 c^2$). 
Recently, BaBar and Belle~\cite{Aubert:2008ju_Wei:2009zv_APS} measured an $\afb$ in the $\bzkstll$ decay larger than the SM expectation.
The decay $\bsphimmpdg$ has not been seen
in previous searches by CDF~\cite{Aaltonen:2008xf_APS} and D0~\cite{Abazov:2006qm_APS}.

In this Letter we report an update of our previous analysis~\cite{Aaltonen:2008xf_APS} of the rare decay modes
$\bpkmm$, $\bzkstmm$, and $\bsphimm$ using an increased data sample of
$p\bar{p}$ collisions at a center-of-mass energy of $\sqrt{s}=1.96~{\rm TeV}$
corresponding to an integrated luminosity of $\lumi$,
collected with the CDF II detector between March 2002 and January 2009.
We update the $\br$ measurements and also report the measurement of
$\afb$ in the $\bzkstmm$ decay.

We reconstruct $\bhmm$ candidates, where $B$ stands for $\bp$, $\bz$, or $\bs$, and
$h$ stands for $K^+$, $\kst$, or $\phi$, respectively.
Charge-conjugation is implied throughout the Letter.
The $\kst$ ($\phi$) meson is reconstructed in the decay $\kst \to K^+\pi^-$ ($\phi \to K^+K^-$).
We also reconstruct $\bpsih$ decays as normalization channels in $\br$ measurements,
because they have final states identical to those of the signals, resulting in a cancellation of 
many systematic uncertainties.
The relative $\br$'s are:
\begin{equation}
 \frac{\br(\bhmm)}{\br(\bpsih)}=
\frac{N_{\hmm}}{N_{\psih}} 
\frac{\varepsilon_{\psih}}{\varepsilon_{\hmm}} \times \br(\psimm),
\label{eq:br}
\end{equation}
 where $N_{\hmm}$ ($N_{\psih}$) is the $\bhmm$ ($\bpsih$) yield, and
${\varepsilon_{\hmm}}/{\varepsilon_{\psih}}$
is the relative reconstruction efficiency determined from the simulation.

The CDF II detector is described in detail in Ref.~\cite{Acosta:2004yw_APS} with the detector subsystems
relevant for this analysis discussed in Ref.~\cite{Aaltonen:2009tz_ref}.

A sample of dimuon events is selected by the online trigger system.
The trigger requires two opposite charged particles with transverse momentum
$p_T\geq1.5$ or $2.0\,\GeVc$ depending on the trigger condition, matched to the muon chambers.
We use the muon chambers detect muons within $|\eta| < 0.6$ and $0.6 < |\eta| < 1.0$~\cite{CDF_coordinate}.
The trigger also requires $L_{xy}>200~\mu \rm m$, where $L_{xy}$ is the transverse displacement of their intersection from the beamline.
The detail of the trigger system and selection requirements can be found in Ref.~\cite{Aaltonen:2008xf_APS}.

The offline loose event selection begins by looking for 
a common vertex of two trigger muons with one (two opposite-charge) reconstructed charged particle(s) 
to form a $\bpkmm$ (a $\bzkstmm$ or a \break $\bsphimm$ ) candidate.
The probability of the vertex fit $\chi^2$ is required to be greater than $10^{-3}$.
All charged particle trajectories are required 
to be associated with hits in the silicon
vertex detector and to  \break have $p_T\geq 0.4\,\GeVc$. In addition, we require \break $p_T(h)\geq 1.0\,\GeVc$
and $p_T(B)\geq 4.0\,\GeVc$.
We require that the $B$ candidate's decay is consistent with being displaced from the primary interaction point in the transverse plane by
$L_{xy}(B)/\sigma(L_{xy}(B))\geq 3$,
and $\sigma(L_{xy}(B))$ is the estimated uncertainty of $L_{xy}(B)$.
We also require that the $B$ candidate comes from the primary vertex by $|d_0(B)|\leq 120~\mu\rm m$, where $d_0(B)$ is the distance of closest approach of the $B$ trajectory to the beamline.

For $\bz$ ($\bs$) candidates the $K^+\pi^-$ ($K^+K^-$) mass must lie within $50$ ($10$)$\,\MeVcsq$ of the world average $\kst$ ($\phi$) mass~\cite{Nakamura:2010zzi_APS}.
The ambiguity of the mass assignment in the $\kst \to K^+\pi^-$ decay is handled by choosing 
the combination with the $K^+\pi^-$ mass closer to the known $\kst$ mass.
This results in the correct mass assignments for about $92\%$ of the decays as determined from the simulation.
Particle identification is performed with the time-of-flight and the ionization energy loss (dE/dx) probabilities of the particle hypothesis.
We require loose particle identification for both kaons and pions coming from the $\kst$ meson or $\phi$ meson to reduce combinatorial background.
This removes $15\%$ of the $B$ mass sideband events while $99.5\%$ of the signal is retained.
We also require a muon likelihood~\cite{Aaltonen:2007kv} to suppress hadron tracks that produce false trigger muons.

Rare decay candidates with a dimuon mass near the $\jpsi$ ($\psi'$) are rejected:
$8.68$ $(12.86)$ $<q^2<$ $10.09$ $(14.18)$ $\,\GeV^2/c^2$.
To eliminate the radiative charmonium decays that escaped rejection above,
we remove candidates consistent with originating from a $B\to \jpsi(')h$ decay followed
by the decay of the $\jpsi(')$ into two muons and a photon: \break
$|(M({\rm \mu\mu h})-M_{B}^{\rm PDG})-(M({\rm \mu\mu})-M_{\rm \jpsi(')}^{\rm PDG})|<100\,\MeVcsq,$
where the PDG superscript indicates known experimental averages~\cite{Nakamura:2010zzi_APS} and
$M(\rm \mu\mu)<M_{\rm \jpsi(')}^{\rm PDG}$.
We also reject candidates if an opposite sign hadron-muon combination of the daughters,
assigned the muon mass, satisfy $\jpsi$ or $\psip$ mass within $40\,\MeVcsq$.
This removes charmonium decays where one of the muons is misidentified as a hadron.
We reject candidates in which two (three) track combinations are compatible within
$\pm25\,\MeVcsq$ with $D^0\to K^-\pi^+$ ($D^+\to K^-\pi^+\pi^+$ or $D_s^+\to K^+K^-\pi^+$) decays for $\bp$, $\bz$, and $\bs$ decays, respectively.
This removes 
$B\to D\pi$ ($D=D^0$, $D^+$, and $D_s^+$) decays where two hadrons are misidentified as muons.

We train an Artificial Neural Network (NN) classifier on simulated signal
and a sample of events representative of the background events in the signal region.
To simulate the signal we use \pythia and \evtgen~\cite{Sjostrand:2000wi_Lange:2001uf_APS} based on the
SM expectation~\cite{Ali:1999mm_APS}.
The background sample is obtained from the
sidebands of the $B$ invariant mass distribution.
We take only the higher mass sideband for the $\bp$ and $\bz$ decays
since the lower sideband is populated with physics background from
partially reconstructed $B$ meson decays.
We use both sidebands for $\bs$ decays.
We use $7$--$10$ observables based on $B$ and daughter's kinematics (e.g. $p_T$ and mass), vertex qualities, and muon likelihoods.
We optimize the NN threshold in order to maximize both the $\br$ and the $\afb$ significance.
For the $\bp$ and $\bz$ analysis we optimize the NN threshold by maximizing $N_s/\sqrt{N_s+N_b}$, 
where $N_s$ ($N_b$) is the expected number of signal (background) events.
We determine $N_s$ by Eq.~(\ref{eq:br}) with the world average $\br$ and NN cut efficiency of the simulated signal events,
and determine $N_b$ from the number of sideband events scaled to the signal region, which is defined as $\pm 2\sigma$ from the world average $B$ mass, and NN cut efficiency of the sideband events.
For $\bs$ decays, $N_s$ is taken from a theoretical prediction~\cite{Geng:2003su_APS}.
We maximize $N_s/(5/2+\sqrt{N_b})$~\cite{Punzi:2003bu_PRL}.

The signal yield is obtained by an unbinned maximum log-likelihood fit to the $B$ candidate invariant mass distribution.
The likelihood is given by \break
${\cal L}=\prod(f_{\rm sig}{\cal P}_{\rm sig}+(1-f_{\rm sig}){\cal P}_{\rm bg})$,
where $f_{\rm sig}$ is the signal fraction,
${\cal P}_{\rm sig}$ is the signal probability density functions (PDF) parametrized with two Gaussian distributions with different means,
and ${\cal P}_{\rm bg}$ is the background PDF modeled with a first- or second-order polynomial.
The signal PDF's are determined from the simulated signal and
the $B$ mass resolution is scaled by the ratio of the mass resolution in $\psih$ data and simulation,
which ranges from 1.07 to 1.09.
The background PDF's are determined from sideband data.
Fitted parameters are $f_{\rm sig}$, the mean $B$ mass, and the background shape.
The fit range for $\bp$ and $\bz$ ($\bs$) decays is from $5.18$ ($5.00$) to $5.70\,\GeVcsq$,
to avoid the region dominated by the physics background.

While the contribution from charmless $B$ decays is negligible due to the muon identification,
we find a sizeable crosstalk between $\bzkstmm$ and \break $\bsphimm$
contributing approximately $1\%$ of the signal, as estimated from simulation.
These contributions, whose fractions are determined by simulation assuming the world average $\br$ and the theoretical prediction~\cite{Geng:2003su_APS}, are subtracted from the fit results for the signal yields.

\begin{figure}[t]
  \begin{center}
    \begin{tabular}{cc}
      \resizebox{.24\textwidth}{!}{\includegraphics[scale=1.00,clip]{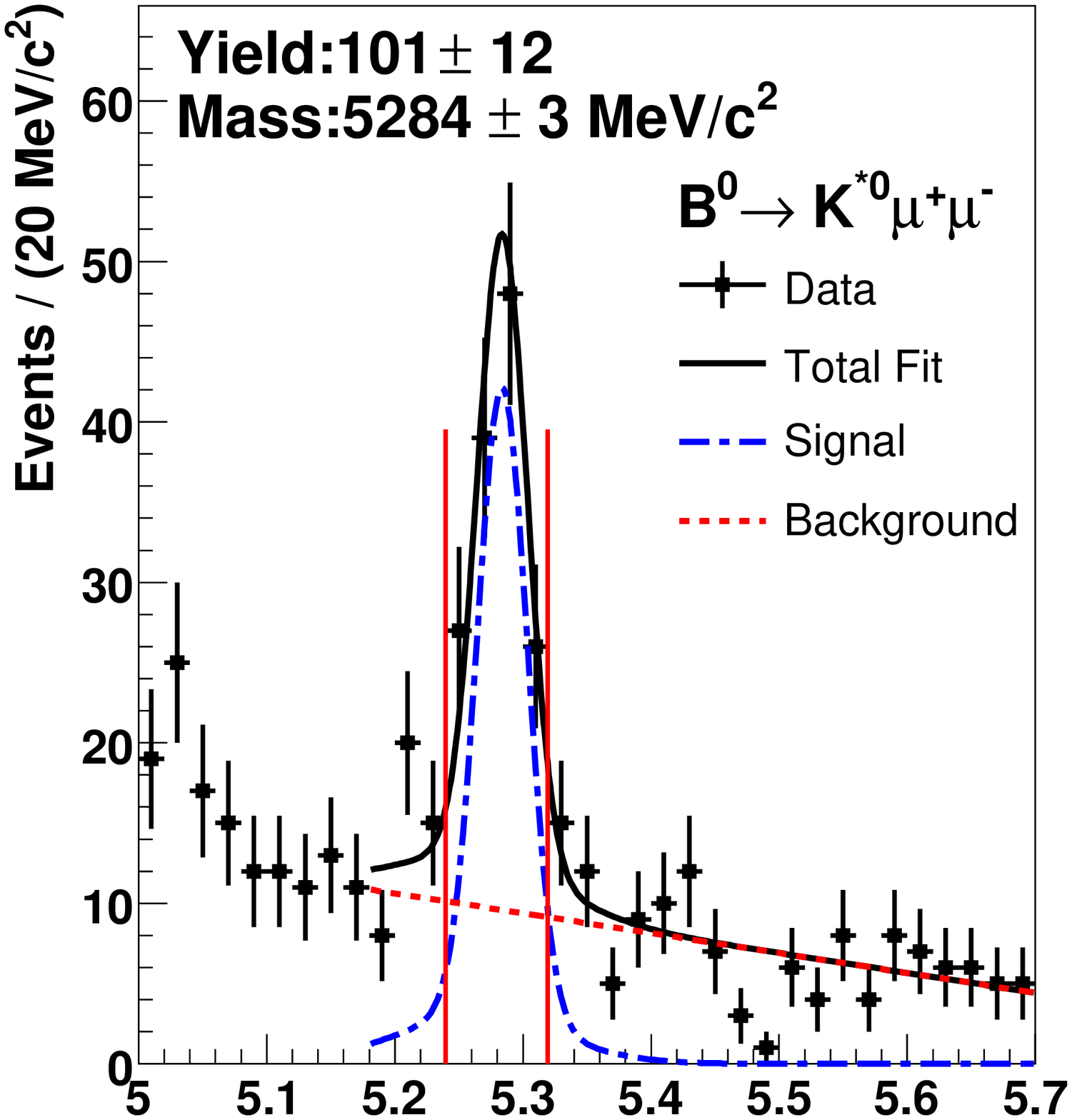}} &
      \resizebox{.24\textwidth}{!}{\includegraphics[scale=1.00,clip]{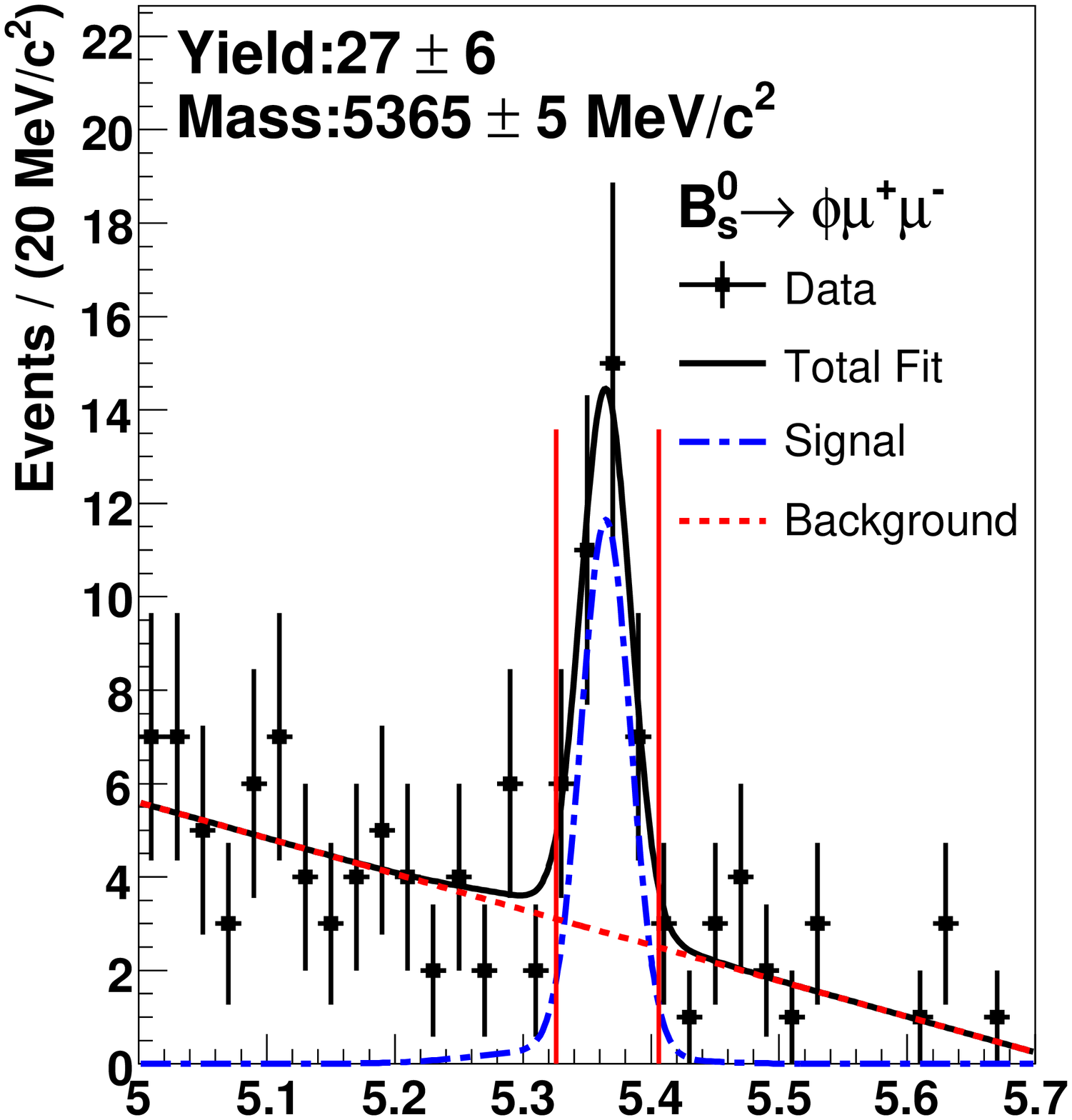}} \\
    \end{tabular}
  \end{center}
  \caption{
Mass of $\bzkstmm$ and $\bsphimm$ candidates with fit results overlaid.
  The vertical lines show the signal region.}
  \label{fig:rarebmass_data}
\end{figure}

By optimized NN threshold we reject $99.5$--$99.8\%$ of background events in the signal region.
Figure~\ref{fig:rarebmass_data} shows the $B$ mass distributions.
The statistical significance is \break
$
s \equiv \sqrt{-2\ln ({\cal L_{\rm null}}/{\cal L_{\rm max}})}
$,
where $\cal L_{\rm max}$ is obtained from a fit with the signal fraction free to float and the mean $B$
meson mass fixed to the fitted value in the corresponding normalization channel, 
and $\cal L_{\rm null}$ is the maximum likelihood obtained from a fit with $f_{\rm sig}=0$.
Systematic uncertainty is not considered in the significance evaluation.
We obtain $s=8.5\sigma$, $9.7\sigma$, and $6.3\sigma$ for $\bp$, $\bz$, and $\bs$ decays, respectively.
The observed yields are listed in Table~\ref{tab:summary_yield}.
This is the first observation of the $\bsphimm$ mode.
\begin{table}[htpb]
  \caption{
    Summary of observed yields.
The numbers in parentheses are the number of events in the signal region.}
  \begin{center}
    \begin{tabular}{cccc}
      \hline\hline
      Mode       & $N_{\hmm}$      & $N_{\psih}$ & ${\varepsilon_{\hmm}}/{\varepsilon_{\psih}}$ \\
      \hline
      $\bp$   & $121\pm16$ ($218$) & $43704\pm245$ ($55296$) & $0.434\pm0.006$ \\
      $\bz$ & $101\pm12$ ($140$) & $15815\pm178$ ($22952$) & $0.477\pm0.009$ \\
      $\bs$ & $27\pm6$ ($40$)  & $2930\pm64$ ($3883$) & $0.498\pm0.012$ \\
       \hline\hline 
    \end{tabular}
  \end{center}
  \label{tab:summary_yield}
\end{table}

 We do not apply a NN selection to $\psih$ channels, because these signals are of sufficient size and purity with the loose selection. 
To obtain the relative efficiency of Eq.~(\ref{eq:br}), the NN cut efficiency of the loosely selected events is considered in addition to the relative efficiency of the loose selection.

The dominant source of systematic uncertainty for each $\br$ measurement is
the background PDF parameterization ($3.9\%$) for $\bp$,
the discrepancy of the NN cut efficiency between data and simulation ($4.8\%$) for $\bz$,
and 
particle identification ($3.5\%$) for $\bs$.
For the absolute $\br$ measurements we assign the uncertainties of the world average $\br(\bpsih)$~\cite{Nakamura:2010zzi_APS}.

Results of the relative $\br$ (Eq.~(\ref{eq:br})) measurements are listed in Table~\ref{tab:summary_br}.
The BR statistical uncertainties include the Poisson term from finite statistics of the sample.
We also show the absolute $\br$ which is obtained by replacing the normalization channel's $\br$ with the corresponding world average~\cite{Nakamura:2010zzi_APS} value.\begin{table}[htpb]
  \caption{ Measured branching fractions of rare modes. First (second) uncertainty is statistical (systematic).} 
  \begin{center}
    \begin{tabular}{lccc}
      \hline\hline
      Mode  & Relative $\br(10^{-3})$ & Absolute $\br(10^{-6})$ \\ 
      \hline
      $\bpkmm$ & $\kmmRelBr \pm \kmmRelBrStat \pm \kmmRelBrSyst$       & $\kmmBr \pm \kmmBrStat \pm \kmmBrSyst$ \\
      $\bzkstmm$ & $\kstmmRelBr \pm \kstmmRelBrStat \pm \kstmmRelBrSyst$ & $\kstmmBr \pm \kstmmBrStat \pm \kstmmBrSyst$ \\
      $\bsphimm$ & $\phimmRelBr \pm \phimmRelBrStat \pm \phimmRelBrSyst$ & $\phimmBr \pm \phimmBrStat \pm \phimmBrSyst$ \\
       \hline\hline 
    \end{tabular}
  \end{center}
  \label{tab:summary_br}
\end{table}

\begin{figure}[t]
  \begin{center}
\setlength{\tabcolsep}{1pt}
\begin{tabular}{cc}
\includegraphics[scale=0.22,clip]{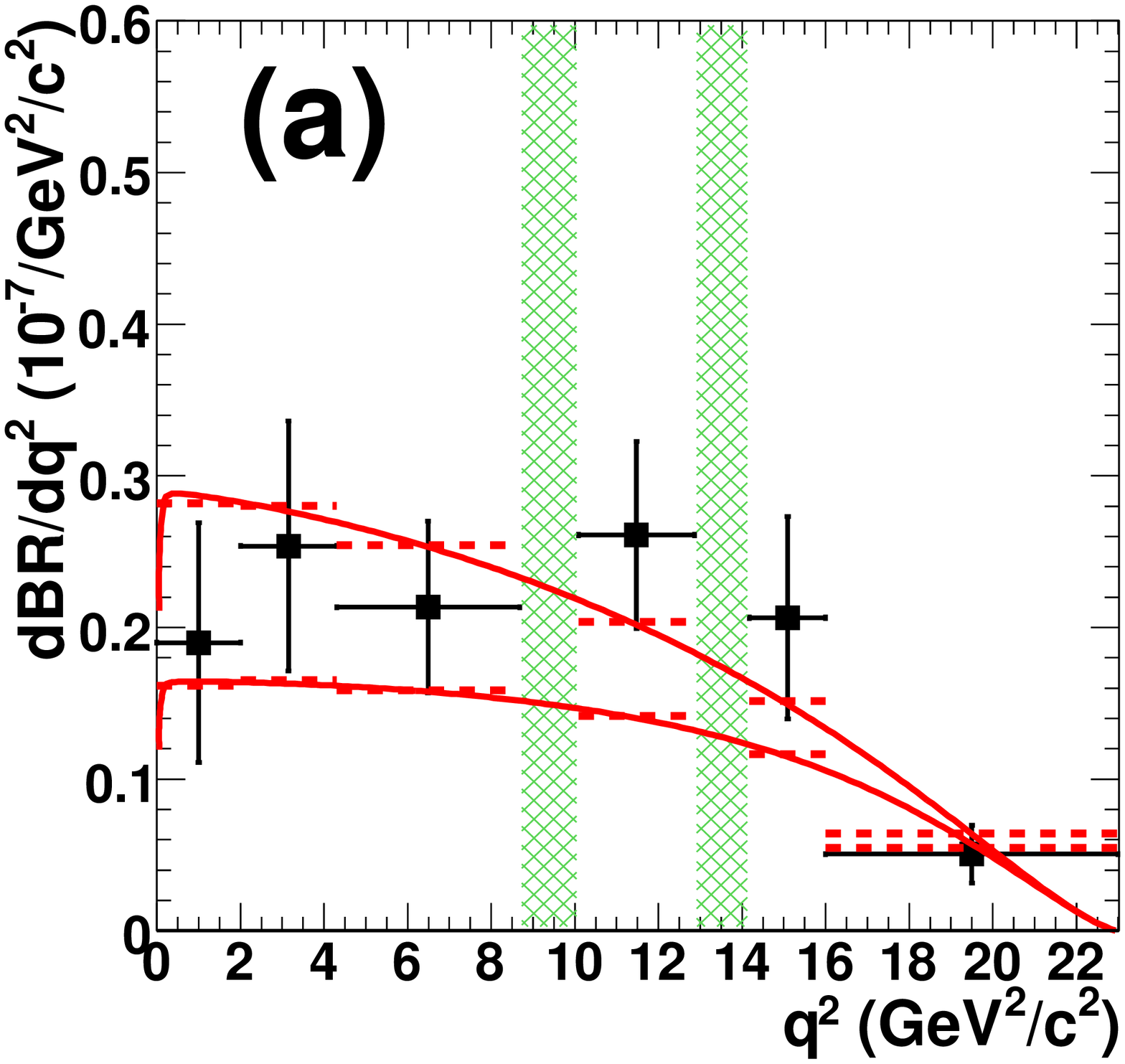} &
\includegraphics[scale=0.22,clip]{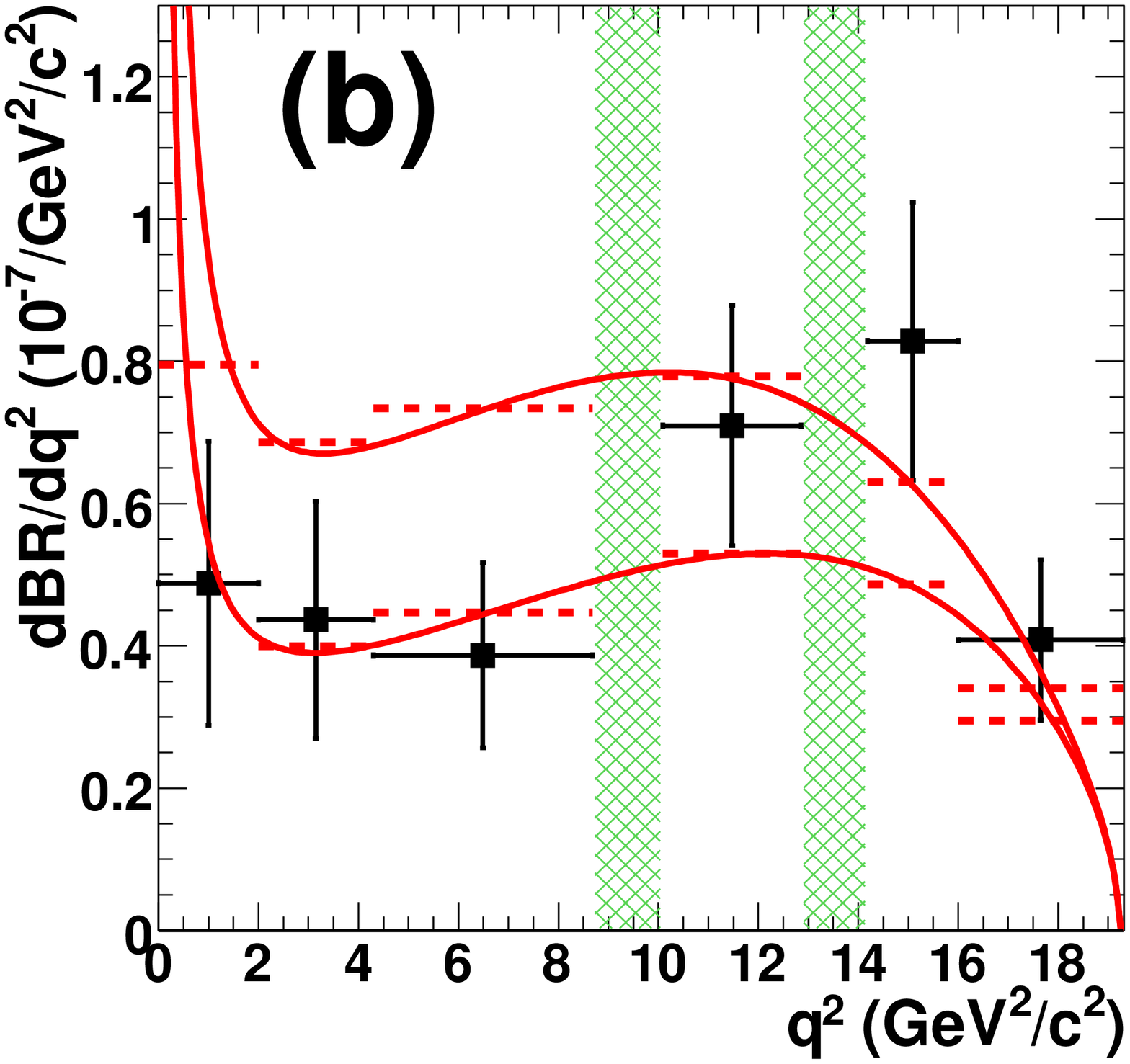} \\
\includegraphics[scale=0.245,clip]{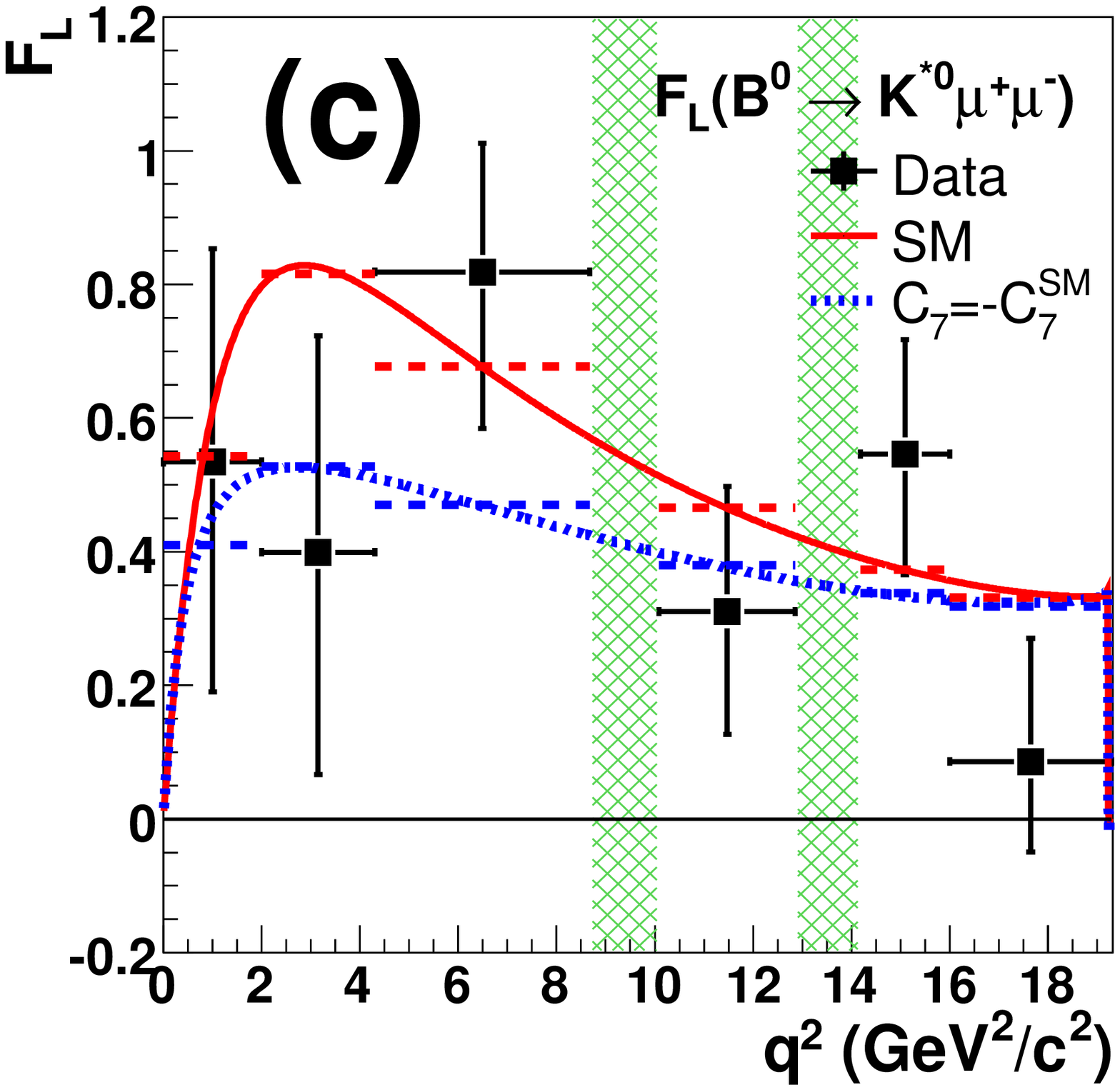} &
\includegraphics[scale=0.245,clip]{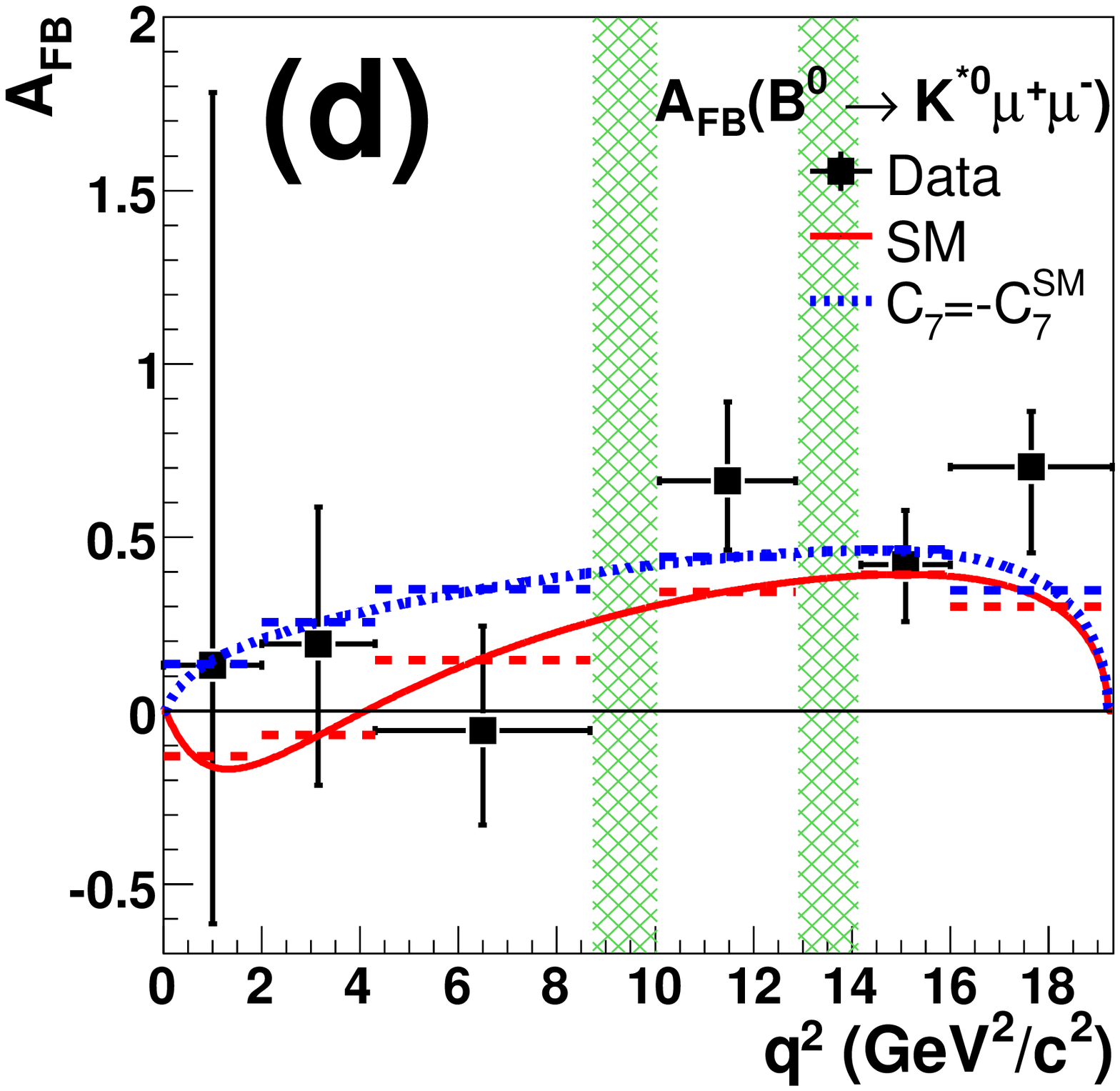} \\
    \end{tabular}
  \end{center}
  \caption{
 Differential $\br$ of $\bpkmm$ (a) and differential $\br$ (b), longitudinal $\kst$ polarization fraction (c), and forward-backward asymmetry (d) of $\bzkstmm$, as a function of squared dimuon mass.
Points are the fit result.
The solid curves are the SM expectation~\cite{Ali:1999mm_APS}.
Two solid curves in (a, b) use maximum- and minimum- \ allowed form factors on differential $\br$ plots.
The dotted curves are the $C_7=-C_7^{\rm SM}$ expectation.
The dashed line is the averaged expectation in each squared dimuon mass bin and hatched regions are charmonium veto regions.}
  \label{fig:dbr_fl_afb}
\end{figure}

These numbers are consistent with our previous results~\cite{Aaltonen:2008xf_APS}, B-factory measurements~\cite{Aubert:2008ps,Aubert:2008ju_Wei:2009zv_APS}, and theoretical expectations~\cite{Geng:2003su_APS}.
We also measure differential $\br$s with respect to the dimuon mass.
Events in the signal mass region are grouped into independent $q^2$ bins.
 Figure~\ref{fig:dbr_fl_afb} (a, b) shows the differential $\br$ for 
$\bpkmm$ and $\bzkstmm$.

The $\afb$ and the $\kst$ longitudinal polarization fraction ($\fl$) are extracted by
an unbinned likelihood fit to the $\cosm$ and $\cosk$ distributions, respectively, where
$\theta_\mu$ is the angle between the $\mu^+$ ($\mu^-$) direction and the direction opposite to the $B$ ($\overline{B}$) meson in the dimuon restframe, and
$\theta_K$ is the angle between the kaon direction and the direction opposite to the $B$ meson in the $\kst$ rest frame.
The differential decay rates~\cite{Kruger:2005ep_APS} are sensitive to $\cosk$ and $\cosm$ through the angular distributions given by 
$
\frac{3}{2}\fl \cosksq + \frac{3}{4}(1-\fl)(1-\cosksq)
$ for $\cosk$ and
$
\frac{3}{4}\fl (1-\cosmsq) + \frac{3}{8}(1-\fl)(1+\cosmsq) + \afb \cosm
$ for $\cosm$. 
We measure $\fl$ and $\afb$ for $\bzkstmm$ and also $\afb$ for $\bpkmm$.
Angular acceptances are obtained from simulated signal samples assuming unpolarized decays. 

The contribution from decays with
$K$-$\pi$ swapped $\kst$ mesons distorts the signal distribution and swaps the sign of $\cosm$.
This effect is considered by adding an additional signal-like term to the likelihood function.
The contribution from decays with non-resonant $K$-$\pi$ is considered to be small~\cite{Kruger:2005ep_APS}
and neglected in the fit.
For the $\bp$ decay, we set $\fl=1$ and consider no scalar term~\cite{Bobeth:2007dw_APS}.

The combinatorial background PDF shape is taken from the $B$ mass upper sideband that is used for the NN training.
In the fit to $\cosk$ ($\cosm$) distribution, the only free parameter is $\fl$ ($\afb$).
For the $\cosm$ fit, the value of $\fl$ is fixed to the $\cosk$ fit result.

Most dominant source of systematic uncertainty for each angular fit is
the fit bias near the physical boundary ($0.02$-$0.07$) for $\fl$ in $\bz$,
the uncertainty of the $\fl$ fit ($0.02$--$0.12$) for $\afb$ in $\bz$,
and
the angular background shape ($0.01$--$0.07$) for $\afb$ in $\bp$.
The total systematic uncertainties lie in the range $0.02$--$0.08$ for $\fl$ in $\bz$,
$0.05$--$0.25$ for $\afb$ in $\bz$, and $0.02$--$0.08$ for $\afb$ in $\bp$.
The angular fit results are shown in Fig.~\ref{fig:dbr_fl_afb} (c, d)
and summarized in Table~\ref{tab:summary_afb}.
Results in the range $0\leq q^2<4.3\,\GeV^2/c^2$ and $1\leq q^2<6\,\GeV^2/c^2$ are also included.

\begin{table*}[htpb]
  \caption{Summary of $\bzkstmm$ and $\bpkmm$ fit results. Maximum $q^2$ is 19.30 (23.00) $\GeV^2/c^2$ for $\bz$ ($\bp$).}
  \begin{center}
    \begin{tabular}{c|ccccc|ccc}
      \hline\hline
      $q^2$ ($\GeV^2/c^2$)   & ${\br}(\bz)\ (10^{-7})$ &{\ \ }& $\fl(\bz)$ &{\ \ }& $\afb(\bz)$ & ${\br}(\bp)\ (10^{-7})$ &{\ \ }& $\afb(\bp)$ \\
      \hline
      $[0.00,2.00)$ & $0.98\pm 0.40\pm 0.09$ && $0.53^{+0.32}_{-0.34}\pm 0.07$ && $\phantom{+}0.13^{+1.65}_{-0.75}\pm 0.25$ & $0.38\pm 0.16\pm 0.03$ && $-0.15^{+0.46}_{-0.39}\pm 0.08$ \\
      $[2.00,4.30)$ & $1.00\pm 0.38\pm 0.09$ && $0.40^{+0.32}_{-0.33}\pm 0.08$ && $\phantom{+}0.19^{+0.40}_{-0.41}\pm 0.14$ & $0.58\pm 0.19\pm 0.04$ && $\phantom{+}0.72^{+0.40}_{-0.35}\pm 0.07$ \\ 
      $[4.30,8.68)$ & $1.69\pm 0.57\pm 0.15$ && $0.82^{+0.19}_{-0.23}\pm 0.07$ && $-0.06^{+0.30}_{-0.28}\pm 0.05$ & $0.93\pm 0.25\pm 0.06$ && $-0.20^{+0.17}_{-0.28}\pm 0.03$ \\ 
      $[10.09,12.86)$ & $1.97\pm 0.47\pm 0.17$ && $0.31^{+0.19}_{-0.18}\pm 0.02$ && $\phantom{+}0.66^{+0.23}_{-0.20}\pm 0.07$ & $0.72\pm 0.17\pm 0.05$ && $-0.10^{+0.17}_{-0.15}\pm 0.07$ \\ 
      $[14.18,16.00)$ & $1.51\pm 0.36\pm 0.13$ && $0.55^{+0.17}_{-0.18}\pm 0.02$ && $\phantom{+}0.42^{+0.16}_{-0.16}\pm 0.09$ & $0.38\pm 0.12\pm 0.03$ && $\phantom{+}0.03^{+0.49}_{-0.16}\pm 0.04$ \\ 
      $[16.00,19.30(23.00))$ & $1.35\pm 0.37\pm 0.12$ && $0.09^{+0.18}_{-0.14}\pm 0.03$ && $\phantom{+}0.70^{+0.16}_{-0.25}\pm 0.10$ & $0.35\pm 0.13\pm 0.02$ && $\phantom{+}0.07^{+0.30}_{-0.23}\pm 0.02$ \\
      \hline
      $[0.00,4.30)$ & $1.98\pm 0.55\pm 0.18$ && $0.47^{+0.23}_{-0.24}\pm 0.03$ && $\phantom{+}0.21^{+0.31}_{-0.33}\pm 0.05$ & $0.96\pm 0.25\pm 0.06$ && $\phantom{+}0.36^{+0.24}_{-0.26}\pm 0.06$ \\
      \hline    
      $[1.00,6.00)$ & $1.60\pm 0.54\pm 0.14$ && $0.50^{+0.27}_{-0.30}\pm 0.03$ && $\phantom{+}0.43^{+0.36}_{-0.37}\pm 0.06$ & $1.01\pm 0.26\pm 0.07$ && $\phantom{+}0.08^{+0.27}_{-0.22}\pm 0.07$ \\
      \hline\hline    
    \end{tabular}
  \end{center}
  \label{tab:summary_afb}
\end{table*}

In summary,
we have updated our previous analysis of the flavor-changing neutral current decays $\bsmm$ using 
data corresponding to an integrated luminosity of $\lumi$.
We report the first observation of the $\bsphimm$, the most rare $\bs$ decay observed to date, and measure the total $\br$.
We measure the total $\br$, differential $\br$, $\afb$ of the $\bpkmm$ and $\bzkstmm$, with respect to $q^2$. 
We also measure $\fl$ of $\bzkstmm$ prior to $\afb$.
These are consistent and competitive with the other current best results.
At present there is no evidence of discrepancy from the SM prediction.

\begin{acknowledgments}
We thank the Fermilab staff and the technical staffs of the participating institutions for their vital contributions. This work was supported by the U.S. Department of Energy and National Science Foundation; the Italian Istituto Nazionale di Fisica Nucleare; the Ministry of Education, Culture, Sports, Science and Technology of Japan; the Natural Sciences and Engineering Research Council of Canada; the National Science Council of the Republic of China; the Swiss National Science Foundation; the A.P. Sloan Foundation; the Bundesministerium f\"ur Bildung und Forschung, Germany; the World Class University Program, the National Research Foundation of Korea; the Science and Technology Facilities Council and the Royal Society, UK; the Institut National de Physique Nucleaire et Physique des Particules/CNRS; the Russian Foundation for Basic Research; the Ministerio de Ciencia e Innovaci\'{o}n, and Programa Consolider-Ingenio 2010, Spain; the Slovak R\&D Agency; and the Academy of Finland. 
\end{acknowledgments}

\bibliographystyle{h-physrev5_cdf_prl}
\bibliography{reference}

\begin{thebibliography}{10}
\bibitem{CDF_coordinate}
We use a cylindrical coordinate system in which $\theta$ is the polar angles
  with respect to the proton beamline and pseudorapidity $\eta \equiv -\ln(\tan
  \theta/2)$.
\end{thebibliography}

\end{document}